\documentclass[a4paper,aps]{revtex4}
\usepackage{graphicx}

\newcommand{\be}{\begin{equation}}
\newcommand{\ee}{\end{equation}}
\newcommand{\ba}{\begin{array}}
\newcommand{\ea}{\end{array}}
\newcommand{\bea}{\begin{eqnarray}}
\newcommand{\eea}{\end{eqnarray}}

\begin{document}

\title{Driven Random Field Ising Model: some exactly solved examples in 
threshold activated kinetics }
\author{Prabodh Shukla} \email{shukla@nehu.ac.in} \affiliation{ Physics 
Department \\ North Eastern Hill University \\Shillong-793 022,India.} 

\begin{abstract} 
The random field Ising Model (RFIM) driven by a slowly varying uniform
external field at zero temperature provides a caricature of several
threshold activated systems. In this model, the non-equilibrium response
of the system can be obtained analytically in one dimension as well as on
a Bethe lattice if the initial state of the system has all spins aligned
parallel to each other. We consider ferromagnetic as well as
anti-ferromagnetic interactions. The ferromagnetic model exhibits
avalanches and non-equilibrium critical behavior. The anti-ferromagnetic
model is marked by the absence of these features. The ferromagnetic model
is Abelian, and the anti-ferromagnetic model is non-Abelian. Theoretical
approaches based on the probabilistic method are discussed in the two
cases, and illustrated by deriving some basic results.
\end{abstract}

\maketitle

\section{Introduction} 

A wide variety of physical systems exhibit threshold activated dynamics,
i.e. they yield to an applied force only if the force has reached a
threshold. For example, a book can be pushed across a table only if the
applied force exceeds the threshold of static friction. More complex
examples involve an infinite array of thresholds. Here we take up an
example from the field of magnetism. A commercial magnet is typically a
two phase material with fine particles of one phase embedded in the other
phase. The precipitation is carried out in a magnetic field, and the
particles are oriented with their long axis parallel to the field
direction. After the material is set in a solid matrix, the needle like
magnetic particles can not orient freely. Instead, the magnetic domains
have to re-orient themselves if the external field is reversed. The
orientation of magnetic domains in fine elongated particles requires a
threshold field. The magnetization of the material in a smoothly
increasing applied field may look smooth on a macroscopic scale but on a
microscopic scale, it is made of steps of irregular widths and heights.
This is known as Barkhausen noise. Thus, the trajectory of magnetization
in an applied field is determined by a threshold activated process. When
the applied field reaches a threshold, the magnetization jumps in a
vertical step. Between two neighboring thresholds of the applied field,
the magnetization remains constant. This marks the width of a step. The
heights and widths of neighboring steps look apparently irregular, but the
spectrum of the Barkhausen noise on a full hysteresis loop shows a
power-law structure. The zero temperature dynamics of the ferromagnetic
random field Ising model ~\cite{sethna} provides a good caricature of this
noise. The model has been solved exactly on a Bethe lattice, and the
power-laws of the Barkhausen noise as well as some other features of the
hysteresis loop have been obtained analytically
\cite{shukla1,dhar,sabha1,shukla2,shukla3,sabha2}.  In the following we
describe the model, and illustrate the method of its solution by
calculating the hysteresis loop as the applied field is cycled from
$-\infty$ to $\infty$ and back. Our method is applicable only if the
initial state has all spins parallel to each other. Therefore, we start
with the applied field equal to $-\infty$ when all spins are aligned along
with the field. One can also calculate minor hysteresis loops within major
hysteresis loop by reversing the applied field at arbitrary points on the
major loop. It is also possible to calculate the distribution of
avalanches (Barkhausen noise) on any part of the hysteresis loop. However,
we do not go into these details here, and refer the reader to the
literature.

We shall also discuss hysteresis in the anti-ferromagnetic random field
Ising model at zero temperature ~\cite{shukla4}.  Hysteresis in the
anti-ferromagnetic RFIM is qualitatively different from that in the
ferromagnetic RFIM because it does not show Barkhausen noise.  On account
of the anti-ferromagnetic interactions, a spin turning up in an increasing
field blocks its nearest neighbors from turning up.  Thus there is no
avalanche of up-turned spins.  Irreversibility or hysteresis in the
ferromagnetic model comes from avalanches. A small increment in the
applied field causes several spins to turn up in the ferromagnetic model,
but reversing the increment does not cause all the spins to turn down. In
the anti-ferromagnetic model, there are no avalanches. So we might think
that there should be no hysteresis in the anti-ferromagnetic model. It is
true that if only one spin turned up at each threshold, there would be no
hysteresis in the anti-ferromagnetic model i.e. the area of the hysteresis
loop would go to zero. However, there is a small amount of hysteresis in
the anti-ferromagnetic model that comes from the following effect. A spin
that turns up, occasionally causes its nearest neighbor (that had turned
up earlier) to turn down. In other words, as the applied field increases
from $-\infty$ to $+\infty$, a small fractions of sites flip three times,
first up, then down, and finally up again.

The lack of avalanches in the anti-ferromagnetic model also sets it apart
from the ferromagnetic model in another aspect. Consider a stable state of
the model in a given applied field, and imagine increasing the applied
field slightly. Suppose there are two neighboring sites that have the
potential to turn up in the increased field. In the ferromagnetic case, we
can turn up any of these spins first, and the other spin would also turn
up subsequently. The exact sequence in which the spins are relaxed in the
ferromagnetic model is not important. This property of the ferromagnetic
model is called the Abelian property.  The anti-ferromagnetic model is
non-Abelian. If two sites can both turn up at a field, it is important to
decide which is to be turned up first i.e. which site has a larger net
field. Once it is turned up the other site is blocked from turning up
until a higher applied field. This constant revision of the threshold
distribution by the dynamics of the anti-ferromagnetic RFIM is the key
difficulty in an exact solution of the model. So far the
anti-ferromagnetic model has been solved exactly in one dimension only.

\section{The Model}

Consider a lattice of N sites. Each site is labeled by an integer i = 1 to
N, and carries an Ising spin $S_{i}$ ( $S_{i}= \pm 1$ ), a quenched random
magnetic field $h_{i}$, and an externally applied uniform field $h$. The
quenched fields $\{ h_{i} \}$ are independent identically distributed
random variables with a continuous probability distribution $\phi(h_{i})$.
There is an interaction J between nearest neighbor spins that can be
ferromagnetic ( J $>$ 0) or anti-ferromagnetic ( J $<$ 0 ). The
Hamiltonian of the system is

\be H=-J \sum_{i,j} S_{i} S_{j} - \sum_{i} h_{i} S_{i} - h \sum_{i} S_{i}
\ee

The spin $S_{i}$ experiences a net field $f_{i}$ on it that is given by,

\be f_{i}=J \sum_{j} S_{j} + h_{i} + h \ee

The Glauber dynamics of the system at temperature T is specified by the
rate $R_{i}$ at which a spin $S_{i}$ flips to $- S_{i}$

\be R_{i}=\frac{1}{\tau} \left[ 1 - S_{i} \mbox{ tanh}\{ f_{i}/(k_{B}T)  
\} \right], \ee

Here $\tau$ sets the basic time scale for the relaxation of individual
spins. The energy of the spin $S_{i}$ is equal to $ - f_{i} S_{i} $. If $
f_{i}$ and $S_{i}$ have the same sign, we say that the spin is aligned
along the net field at its site. The energy of a spin is the lowest if it
is aligned along the net field at its sight. We are interested in the
dynamics of the model at zero temperature ( T = 0 ), and on time scales
much larger than $\tau$. In this limit, the dynamics simplifies to the
following rule: choose a spin at random, and flip it only if it is not
aligned along the net field at its site.  Repeat the process till all
spins are aligned along the net fields at their respective sites.

We apply the above dynamics to calculate hysteresis in the system when it
is driven by a uniform external field $h(t)=-h_{0} \mbox{ cos } \omega t$.
We work in the limit $h_{0} \rightarrow \infty$, and $\omega \rightarrow
0$. In other words, we start with a sufficiently large negative applied
field $h_{0}$ so that the stable configuration has all spins down ( $S_{i}
= -1$, for all $i$), and increase the field slowly. At some value of
$h(t)$, the net field $f_{i}$ at some site $i$ will become positive, and
$S_{i}$ would flip up. This changes the net field at the neighbors, and in
the ferromagnetic case some of the neighbors may flip up, and so on,
causing an avalanche of flipped spins. We keep the applied field fixed
during an avalanche. After the avalanche has stopped and a stable state
has been reached, we calculate the magnetization per site m(h);

\be m(h) = \frac{1}{N} \sum_{i} S_{i} \ee.

Then we raise the applied field slightly till the next avalanche occurs,
and calculate the magnetization again after that avalanche has stopped.
This process is continued until the applied field is sufficiently large and
positive ($h_{0}$), and all spins are up. The magnetization trajectory
$m_{l}(h)$ determined in this way gives the lower half of the hysteresis
loop in the limit of zero driving frequency. The upper half of the
hysteresis loop $m_{u}(h)$ is similarly obtained by decreasing the applied
field from $h_{0}$ to $-h_{0}$. This is related to $m_{l}(h)$ by symmetry,
$m_{u}(h)$ = $-m_{l}(-h)$. Note that the area of the hysteresis loop in
this model does not go to zero even if the driving frequency goes to zero
because the limit T $\rightarrow$ 0 has been taken before the limit
$\omega \rightarrow$ 0.

The hysteresis loop obtained by the zero temperature dynamics of the
random field Ising model described above is a good model of low
temperature hysteresis seen experimentally in several systems. In
experimental systems, the magnetization states on the two halves of the
hysteresis loop are metastable states with a life time much longer than
any other time scale of practical interest. Our model replaces the
metastable states by fixed point states, but the fixed points correspond
to the metastable states in the sense that they represent local minima of
energy at a particular applied field. In this context, it is useful to
consider the time scales of interest in physical systems, and the
appropriateness of the zero-temperature dynamics as a model. In a typical
hysteresis experiment, there are at least four time scales:  (i) typical
time $\tau$ that an individual spin takes to relax, (ii) time $\tau_{1}$
that the system takes to relax to a metastable state, (iii) time
$\tau_{2}$ over which the applied field changes, and (iv) life time
$\tau_{3}$ of the metastable state. In complex systems, $\tau_{1}$,
$\tau_{2}$, and $\tau_{3}$ may each contain an entire spectrum of time
scales. In physical systems relevant to our model, the shortest time is
$\tau$ which may be taken to be unity to set the scale. The next larger
time is $\tau_{1}=\nu \times \tau$, where $\nu$ is the number of
iterations of the dynamics to reach a fixed point. The applied field is
assumed to vary very slowly (driving frequency goes to zero!) so that
$\tau_{2} >> \tau_{1}$. Specifically, it means that the applied field is
held constant during the relaxation of the system. The time $\tau_{3}$ is
infinite. Thus, the present model is applicable if $ \tau_{3} >> \tau_{2}
>> \tau_{1} >> \tau$. These conditions fit a wide class of complex
magnetic materials that have a large number of metastable states separated
from each other by barriers much larger than the available thermal energy.

\section{Ferromagnetic Model on a Bethe Lattice}

A Bethe lattice is an infinite-size branching tree of coordination number
$z$. We choose a site at random ( call it the central site ), and ask for
the probability that this site is up when the applied field has been
increased from $-\infty$ to $h$. Each nearest neighbor of the central site
forms the vertex of an infinite sub-tree. If the central site were to be
deleted, the lattice would break up into $z$ disjointed pieces. This means
that the spin-flip dynamics on $z$ sub-trees is independent of each other
as long as the central site does not flip up from its initial state. The
Abelian nature of the ferromagnetic model allows us to relax the sites in
any order of our choice. We choose to relax the central site after its
neighbors have been relaxed. Let $P^{*}(h)$ denote the conditional
probability that a nearest neighbor of the central site is up at field $h$
before the central site is relaxed (i.e. given that the central site is
down). Then the probability that the central site is up at $h$ is given
by,

\be p(h)=\sum_{n=0}^{z} \left( \ba{c} {z} \\ {n} \ea \right)
[P^{*}(h)]^{n} [1-P^{*}(h)]^{z-n} p_{n}(h) \ee

The last factor, $p_{n}(h)$, gives the probability that a spin with $n$
nearest neighbors up, and $z-n$ neighbors down has sufficient random field
to flip it up at an applied field $h$.

\be p_{n}(h) = \int_{(z - 2n) J -h}^{\infty} \phi(h_{i}) dh_{i} \ee

The magnetization per site is given by $m(h)=2 p(h) -1$. The calculation
of $P^{*}$ proceeds as follows. Consider a nearest neighbor of the central
site and its sub-tree. Let the nearest neighbor i.e. the vertex of the
sub-tree be $n$ steps away from the boundary of the tree (the limit n
$\rightarrow \infty $ will follow). The sites on the boundary of the
sub-tree have only one nearest neighbor situated at a height of one step
from the boundary. We relax the boundary sites first holding sites at
height-1 down.  Then we relax sites at height-1, holding sites at height-2
down., and continue this process. The conditional probability $P^{m}$ that
a site at height-m is up given that its nearest neighbor at height $m+1$
is held down, satisfies the recursion relation,

\be P^{m}(h)=\sum_{n=0}^{z-1} \left( \ba{c} {z-1} \\ {n} \ea \right)
[P^{m-1}(h)]^{n} [1-P^{m-1}(h)]^{z-1-n} p_{n}(h). \ee

The above equation is understood as follows. A site at height $m$ has one
neighbor at height $m+1$, and $z-1$ neighbors at a height $m-1$.  The
neighbor at height $m+1$ is held down while the site at $m$ is relaxed.
Each neighbor at height $m-1$ can be independently up with probability
$P^{m-1}(h)$ before the site at height $m$ is relaxed. The last factor
$p_{n}(h)$ gives the probability that the site at height $m$ flips up when
relaxed. As, $m \rightarrow \infty$, the above equation iterates to the
fixed point value $P^{*}(h)$.

If the applied field is reversed before completing the lower half of the
major hysteresis loop, we generate what is known as a minor hysteresis
loop. First reversal of the field generates the upper half of the minor
loop, and a second reversal generates the lower half.  When the field on
second reversal reaches the point where the first reversal was made, the
lower half of the minor loop meets the starting point of the upper half.
That is, the minor loop closes upon itself at the point it started. This
property of the RFIM is known as return point memory. 

The analytic calculation of minor loop is more difficult technically than
that of the major loop. Consider the upper half of the minor loop. Suppose
the applied field is reversed from $h$ to $h^{\prime}$ ($h^{\prime} \le
h$). We want to know the probability that an arbitrary site $i$ which was
up at $h$ turns down at $h^{\prime}$. It is necessary to know how many
nearest neighbors of site $i$ are up at $h^{\prime}$. But this is not
enough. We also need to know how many of the up neighbors were up before
site $i$ turned up, and how many turned up after site $i$. When site $i$
turns up, the field on each nearest neighbor increases by an amount $2J$.
It does not affect the neighbors which are already up but some down
neighbors may turn up as a result of the increased field. For each down
neighbor that turns up after site $i$, the field on site $i$ increases by
an amount $2J$. Even if one down neighbor turns up, site $i$ will not turn
down if the applied field that causes it to turn up is rolled back
infinitesimally. It has to be rolled back sufficiently so that the
neighbor which turned up after site $i$ turns back down again. Site $i$
can not turn down until all neighbors which turned up after it have turned
down. This is the origin of hysteresis in the model, and also the
difficulty in the calculation of the minor loop. We have to calculate the
probability $D^{*}(h^{\prime})$ that a nearest neighbor of site $i$ that
was down before site $i$ turned up is down again at $h^{\prime}$.
$D^{*}(h^{\prime})$ is determined by the equation,

\bea D^{*}(h^{\prime})= \sum_{n=0}^{z-1} \left( \ba{c} {z-1} \\ {n} \ea
\right) [P^{*}_{l}(h)]^{n}[1 - P^{*}_{l}(h)]^{z-1-n}
\left[1-p_{n+1}(h)\right] & \nonumber \\ +\sum_{n=0}^{z-1} \left( \ba{c}
{z-1} \\ {n} \ea \right) [P^{*}_{l}(h)]^{n}[D^{*}(h^{\prime})]^{z-1-n}
\left[p_{n+1}(h) - p_{n+1}(h^{\prime})\right] \eea

Given a site $i$ that is up at $h$, the first sum above gives the
conditional probability that a nearest neighbor of site $i$ remains down
at $h$ after site $i$ has turned up. The second sum takes into account the
situation that the nearest neighbor in question turns up at $h$ after site
$i$ turns up but turns down at $h^{\prime}$.

The fraction of up sites which turn down at $h^{\prime}$ is given by,
\be
q^{\prime}(h^{\prime})=\sum_{n=0}^{z} \left( \ba{c} {z} \\ {n} \ea \right)
[P^{*}_{l}(h)]^{n}[D^{*}(h^{\prime})]^{z-n}
\left[p_{n}(h)-p_{n}(h^{\prime})\right]
\ee

The magnetization on the upper return loop is given by,
\be
m^{\prime}(h^{\prime})=2 [p(h)-q^{\prime}(h^{\prime})] -1
\ee

We reverse the field $h^{\prime}$ to $h^{\prime\prime}$ ($h^{\prime\prime}
> h^{\prime}$) to trace the lower half of the return loop. The
magnetization on the lower half of the return loop may be written as,

\be
m^{\prime\prime}(h^{\prime\prime})=2 [p(h)-q^{\prime}(h^{\prime})
+ p^{\prime\prime}(h^{\prime\prime})] -1
\ee

where $p^{\prime\prime}(h^{\prime\prime})$ is the probability that
an arbitrary site $i$ which turned up at $h$ and turned down at
$h^{\prime}$, turns up again at $h^{\prime\prime}$.

\be
p^{\prime\prime}(h^{\prime\prime})=\sum_{n=0}^{z} \left( \ba{c} {z}
\\ {n} \ea \right) [U^{*}(h^{\prime\prime})]^{n}
[D^{*}(h^{\prime})]^{z-n}
\left[p_{n}(h^{\prime\prime})-p_{n}(h^{\prime})\right]
\ee

Here $U^{*}(h^{\prime\prime})$ is the conditional probability that
a nearest neighbor of a site $i$ turns up before site $i$ turns up
on the lower return loop. It is determined by the equation,

\bea
U^{*}(h^{\prime\prime})= P^{*}(h) -\sum_{n=0}^{z-1} \left( \ba{c} {z-1}
\\ {n} \ea \right) [P^{*}_{l}(h)]^{n}[D^{*}(h^{\prime})]^{z-1-n}
[p_{n}(h)-p_{n}(h^{\prime})] & \nonumber \\
+\sum_{n=0}^{z-1} \left( \ba{c} {z-1} \\ {n} \ea \right)
[U^{*}(h^{\prime\prime})]^{n}[D^{*}(h^{\prime})]^{z-1-n}
\left[p_{n}(h^{\prime\prime}) - p_{n}(h^{\prime})\right]
\eea

The rationale behind equation (13) is similar to the one behind equation
(8). Given that a site $i$ is down at $h^{\prime}$, the first two terms
account for the probability that a nearest neighbor of site $i$ is up at
$h^{\prime\prime} \ge h^{\prime}$. Note that the neighbor in question must
have been up at $h$ in order to be up at $h^{\prime}$, and if it it is
already up at $h^{\prime}$ then it will remain up on the entire lower half
of the return loop, i.e. at $h^{\prime\prime} \ge h^{\prime}$. The third
term gives the probability that the neighboring site was down at
$h^{\prime}$, but turned up on the lower return loop before site $i$
turned up. It can be verified that the lower return loop meets the lower
major loop at $h^{\prime\prime}=h$ and merges with it for
$h^{\prime\prime} > h$ as may be expected on account of the return point
memory.

The analytic results on the Bethe lattice are obtained by taking the
infinite-size limit of a branching tree. Fixed points of recursion
relations have the effect of eliminating surface effects. It is rather
difficult to eliminate surface effects in numerical calculations on
branching trees because most of the sites on a finite tree are on the
boundary or close to the boundary. We therefore perform the numerical
simulations of the model on random graphs of coordination $z$. A random
graph of $N$ sites has no surface, but the price we pay is that it has
some loops. However, for almost all sites in the graph, the local
connectivity up to a distance of $ \mbox{log}_{(z-1)}$ N is similar to the
one in the deep interior of the branching tree. Therefore, simulation on a
random graph is a very efficient method of subtracting the surface effects
on the corresponding finite branching tree. Figure 1 shows a comparison
between the theoretical expression and the numerical simulation for z=4,
and $\sigma=1.7$. The agreement is quite remarkable considering the
simulations are performed on a random graph, and the theoretical result is
obtained by taking the infinite size limit of a branching tree.

The analysis presented above shows that a reversal of the applied field by
an amount 2J at any point on the lower half of the major hysteresis loop
brings the system on the upper half of the loop. This result provides an
interesting possibility for measuring the exchange interaction J in a
hysteresis experiment. Another point that is brought out by the above
analysis is the following. Hysteresis on a Bethe lattice of coordination z
$>$ 3 is qualitatively different from the case z=2, and z=3. For $ z \ge
4$, there is a critical value of $\sigma$ that characterizes the Gaussian
random field distribution. If $\sigma$ is less than the critical value
$\sigma_{c}$, the magnetization in increasing field has a macroscopic
first-order jump at an applied field $h_{c}$ $>$ J. As $\sigma$ increases
to $\sigma_{c}$, $h_{c}$ decreases to J, and the first-order jump in
magnetization reduces to zero. The system shows non-equilibrium critical
behavior at $h=h_{c}$, and $\sigma=\sigma_{c}$. For z=4, $\sigma_{c}=1.78$
approximately. Bethe lattices with z=2, and z=3 do not show a macroscopic
jump in the magnetization or critical behavior for any value of $\sigma $.
Perhaps this dependence of the non-equilibrium critical behavior on the
coordination number of the lattice has a more general validity going
beyond the Bethe lattice. Numerical simulations and theoretical arguments
on several periodic lattices embedded in two and three-dimensional space
show that hysteresis on periodic lattices with $z \ge 4$ is qualitatively
different from that on lattices with $z < 4$. Although there are some
similarities between Bethe lattices and periodic lattices of the same
coordination number, there are differences as well. The differences are
related to the bootstrap percolation instability on some periodic lattices
~\cite{sabha3}.

The method of calculating the minor loop described above may also be
extended to obtain a series of minor loops nested within the minor loop
obtained above. The key point is that whenever the applied field is
reversed, a site $i$ may flip only after all neighbors of site $i$ which
flipped in the wake of site $i$ (on the immediately preceding sector) have
flipped back. The neighbors of site $i$ which remained firm after site $i$
flipped do not yield before site $i$ has flipped. We have obtained above
expression for the return loop when the applied field is reversed from
$h_{ext}=h$ on the lower major loop to $h_{ext}=h^{\prime}$ ($h-2J \le
h^{\prime} \le h$), and reversed again from $h_{ext}=h^{\prime}$ to
$h_{ext}=h^{\prime\prime}$ ($h^{\prime\prime} \le h$). When the applied
field is reversed a third time from $h^{\prime\prime}$ to
$h^{\prime\prime\prime}$ ($h^{\prime\prime\prime}< h^{\prime\prime}$),
expressions for the magnetization on the nested return loop follow the
same structure as the one on the trajectory from $h$ to $h^{\prime}$.
Qualitatively, the role of $P^{*}$ on the first leg ($h$ to $h^{\prime}$)
is taken up by $U^{*}$ on the third leg ($h^{\prime\prime}$ to
$h^{\prime\prime\prime}$) of the nested return loop.

It was noted earlier ~\cite{shukla4} that the major hysteresis loops for
the RFIM on Bethe lattices of coordination number $z$ have discontinuities
for certain distributions of the random field. For a Gaussian distribution
of the random field with mean value zero and variance ${\sigma}^{2}$, a
discontinuity in each half of the major hysteresis loop occurs for values
of $\sigma$ smaller than a critical value, and for $z \ge 4$. There is no
discontinuity for a Gaussian distribution on lattices with $z=2$, and
$z=3$.  These considerations apply to minor loops as well. The set of
equations determining the minor loops are polynomials of degree $z$, and
stable solutions of these equations in a continuously changing applied
field may change discontinuously. At a critical applied field where the
major loop has a jump discontinuity, we have the possibility of two minor
loops depending upon the magnetization of the state which is used to
generate the minor loop.

\section{Anti-Ferromagnetic Model in One Dimension}

As mentioned earlier, the zero temperature dynamics of the
anti-ferromagnetic random field Ising model is non-Abelian. The stable
configuration in an applied field $h$ depends on the order in which the
sites are relaxed, in addition to the hysteretic dependence on the initial
state of the system. As we raise the applied field from $h=-\infty$ very
slowly, spins turn up in a sequence in which the net fields at their sites
vanish. When a spin turns up, the net field on its neighbors is decreased
by an amount $2|J|$. This blocks the neighbors from turning up until the
applied field has increased by at least an amount $2|J|$.

In our model, the rate at which the applied field is increased does not
have any fundamental significance. We can expose the initial state ( with
all spins down ) directly in the field $h$, and relax the spins in the
order they would have followed if the field were increased infinitely
slowly. Let $p_{0}(h)$ denote the probability that in an applied field
$h$, a spin with random field $h_{i}$ has the potential to turn up if none
of its neighbors are up, i.e. $p_{0}(h)$ is the probability that the
quantity $[h_{i} + 2 |J| + h]$ is greater than zero before the relaxation
process is executed. During relaxation, the fraction of sites that
actually turn up may be much less than $p_{0}(h)$, especially for large
$p_{0}(h)$. The reason is that the potential sites have to be relaxed
sequentially in the order of decreasing net field on them. A potential
site that is adjacent to another potential site with a higher field gets
blocked by it. It takes a relatively larger applied field to turn up a
spin at a blocked site.

We consider the stable state of a one dimensional model in an applied
field $h$. If $h$ is not too high, the stable state may be punctuated by
pairs of adjacent spins along the chain that have never flipped from their
initial state. These pairs of down spins (to be called doublets below)
serve to shield the dynamics of the chain on one side of the pair from
that on the other side. The evolution on a finite segment of the chain
between two doublets is independent of the rest of the chain. It is
therefore of interest to calculate the density of doublets in a stable
chain. If $P_{\downarrow \downarrow} $ denotes the conditional probability
that a spin is down in a stable state given that one of its nearest
neighbors is down, then $P_{\downarrow \downarrow}^{2} $ is the
probability per site of observing a doublet on the chain.

Consider a string of n+1 sites embedded in the chain. Number the first
site at the left end of the string as 0, and the following sites as 1, 2,
\ldots n. Given that site-0 is down, we wish to calculate the conditional
probability $P_{\downarrow \downarrow}[0\downarrow:1\downarrow] $ that
site-1 is down.  Site-1 may be down because it does not have the potential
to turn up, or it may have the potential to turn up but may be blocked by
site-2. If $m$ nearest neighbors of a spin are up just before that spin
turns up, we say that it has turned up under a $p_{m}$-process
($m$=0,1,2). If a spin does not have the potential to turn up under a
$p_{0}$-process, then it does not have the potential to turn up at all.
The probability that site-1 does not have the potential to turn up is
equal to $1 - p_{0}(h)$. Thus,

\be P_{\downarrow \downarrow}[0\downarrow:1\downarrow] = 1 - p_{0}(h);
\mbox{    ( $2|J|+h_{1}+h < 0$ ). } \ee

Calculation of the probability that site-1 has the potential to turn up
but is blocked by site-2 is relatively complicated. Our strategy is to
calculate it in an approximation initially, and then incorporate
corrections in our calculation to make it exact. The approximation we use
is to assume that only $p_{0}$-processes operate initially; $p_{1}$ and
$p_{2}$-processes are put on hold temporarily. In other words, only those
spins turn up whose neighbors on both sides are down. If the spin at
site-1 is blocked, then the spin at site-2 must be up. The probability
that the spin at site-2 is up depends upon how the spins have turned up on
the right of site-2.  How far does the correlation extend? The correlation
must end at the first occurrence of a doublet. But it may end earlier at a
site that is the first site to flip up on the right of site-1. Suppose
site-n is the first site to flip up. Then site-1 will be correlated to
site-n if $h_{n} > h_{n-1} > h_{n-2} \ldots h_{2}> h_{1}$. Note that if
the above chain of inequalities is not satisfied, we would get a doublet
between site-1 and site-n contrary to our assumption. The probability of
getting n potential sites with random fields ordered as above is
$p_{0}^{n}/n!$. In this case, the next site to turn up after site-n will
be site-(n-2) followed by sites (n-4), (n-6), and so on. Site-1 will end
up being down if n is even, and up if n is odd. Therefore the probability
that site-1 will be down at the end of the relaxation process is given by,

\begin{displaymath} P_{\downarrow \downarrow}^{a}[0\downarrow:1\downarrow]
= \sum_{n=2}^{\infty} \frac{[-p_{0}(h)]^{n}}{n!} \mbox{ ( $2|J| +h_{1}+h
>0$; $p_{1},p_{2}$ \mbox{-processes on hold.} ) } \end{displaymath}

We can rewrite the above equation as, \bea P_{\downarrow
\downarrow}^{a}[0\downarrow:1\downarrow] = g(h);( \mbox{$2|J| +h_{1}+h
>0$; $p_{1},p_{2}$-processes on hold.}); \nonumber\\ g(h)= \exp{ \{
-p_{0}(h) \} } - \{1 - p_{0}(h)\} \eea

Combining the preceding two equations, we get the conditional probability
that the spin at site-1 is down whether site-1 is a potential site or not.

\be P_{\downarrow \downarrow}[0\downarrow:1\downarrow] =
\exp{\{-p_{0}(h)\}} \mbox{ ( $p_{1},p_{2}$ \mbox{-processes on hold.} )}\ee

Now we put in the corrections to get an exact result. If $p_{1}$ and
$p_{2}$-processes are allowed to proceed along with the $p_{0}$-process,
site-2 can turn up in two ways. It may turn up under a $p_{0}$-process, or
it may turn up under a $p_{1}$-process. If site-2 turns up under a $p_{0}$
process, site-1 may turn up under a $p_{1}$- process subsequently. This
event decreases $P_{\downarrow \downarrow}[0\downarrow:1\downarrow]$ below
the value $\exp{(-p_{0})}$ calculated above. On the other hand, if site-3
turns up under a $p_{0}$-process , and subsequently if site-2 turns up
under a $p_{1}$-process, site-1 which would have otherwise turned up under
a $p_{0}$-process may stay down on account of being blocked by site-2.
This event increases $P_{\downarrow \downarrow}[0\downarrow:1\downarrow]$
above $\exp{(-p_{0})}$. In the following we calculate the two correction
terms mentioned above.

\begin{enumerate}

\item If the spin site-2 turns up under a $p_{0}$-process, the probability
that the spin at site-1 is down, and $h_{1} +2|J| + h > $ 0, is equal to
g(h). If site-1 is to turn up after site-2, we must have $h_{1} + h >$0.
The probability for this event is $g(h-2|J|)$. The amount $g(h-2|J|)$ has
to be subtracted from $\exp{(-p_{0})}$.

\item Consider the event that the spin at site-3 turns up under a
$p_{0}$-process, subsequently the spin at site-2 turns up under a
$p_{1}$-process, and the spin at site-1 stays down. If the spin at site-1
were to turn up as well, then we do not get any additional contribution to
$P_{\downarrow \downarrow}[0\downarrow:1\downarrow]$. The probability that
the spin at site-1 remains down is given by,

\be P_{\downarrow \downarrow}^{a}[0\downarrow:1\downarrow] =
\int_{-h-2|J|}^{-h} \phi(h_{1})  g(-h_{1}-4|J|)\mbox{ } dh_{1} \ee

The limits on integration take into account the fact that the random field
$h_1$ at site-1 has to satisfy the inequalities $h_1+2|J|+h\ge0$, and
$h_1+h <0$. Also, we must have $h_2 \ge h_1+2|J|$ in order for site-2 to
flip up under a $p_1$-process before site-1 can flip under a
$p_0$-process. The probability for this is given by $g(-h_1-4|J|)$. The
amount $P_{\downarrow \downarrow}^{a}[0\downarrow:1\downarrow]$ is to be
added to $\exp{(-p_{0})}$. 

\end{enumerate}

Putting the three terms together, the conditional probability that site-1
is down given that site-0 is down is,

\be P_{\downarrow \downarrow}[0\downarrow:1\downarrow] =
\exp{\{-p_{0}(h)\}} - g(h-2|J|) + P_{\downarrow
\downarrow}^{a}[0\downarrow:1\downarrow] \ee

Cognizant readers may notice that we have considered the possibility of
only one spin turning up under a $p_{1}$-process. But several spins may
turn up under a $p_{1}$-process. Indeed, the quenched field at site-1 is
the smallest of the set of quenched fields $ \{h_{n} > h_{n-1} > h_{n-2}
\ldots h_{2}> h_{1} \}$. Therefore, if the spin at site-1 can turn up
under a $p_{1}$-process then all other spins under consideration can also
turn up under a $p_{1}$-process. Suppose spin at site-4 turns up under a
$p_{0}$-process, and subsequently spins at site-3, and site-2 turn up
under a $p_{1}$-process. This would happen if $h_{3}-h_{2} > 2|J|$, and
$h_{2}-h_{1} > 2|J|$. However, the resulting state in this case could be
obtained by relaxing site-4 and site-2 under a $p_{0}$-process, and then
site-3 under a $p_{2}$-process. Care has to be exercised when calculating
the fraction of up spins in this way (some spins may turn down when their
neighbor turns up), but it does not contribute any extra term to doublets.
Similarly, the extra doublet that was considered above need not
necessarily occur on site-0 and site-1. It can occur at any other position
between site-1 and site-n, but the correction term $P_{\downarrow
\downarrow}^{a}[0\downarrow:1\downarrow]$ remains the same. The
significance of site-1 in our discussion is only that its left neighbor at
site-0 is held fixed in the down state.

In Figure (2) we show the calculated probability of observing doublets in
a stable state of the anti-ferromagnetic chain at an applied field $h$ for
a Gaussian distribution of the random field with $\sigma$=.5, and
$\sigma$= 1. In each case, result from an appropriate numerical simulation
has been superimposed on the theoretical expression. The agreement between
the theoretical expression and the corresponding numerical result is quite
good. In fact, on the scale of figure (2) the two are indistinguishable
from each other.

The calculation of doublets presented above allows us in principle to
determine the modified a posteriori distribution of random fields on all
spins that are down. We can therefore calculate the probability of any
spin turning up under a $p_{0}$, $p_{1}$, or $p_{2}$-process. Thus the
magnetization at any applied field $h$ may be calculated. The details
however are rather tedious. For a rectangular distribution of the random
field with mean zero and width 2$\Delta$ ($\Delta < |J|$), the reader may
refer to reference ~\cite{shukla4}. The simplifying feature of the case
$\Delta < |J|$ is that ($h_{i} - h_{i-1} < 2|J|$) for any site $i$ on the
chain. Therefore no $p_{1}$-process can take place before all
$p_{0}$-processes have been exhausted. Similarly no $p_{2}$-process can
take place before all $p_{1}$-processes have been exhausted. The details
of the calculation of magnetization for an unbounded distribution of the
random field will be given elsewhere.

\begin{figure}
\begin{center}

\includegraphics[angle=-90,width=16cm]{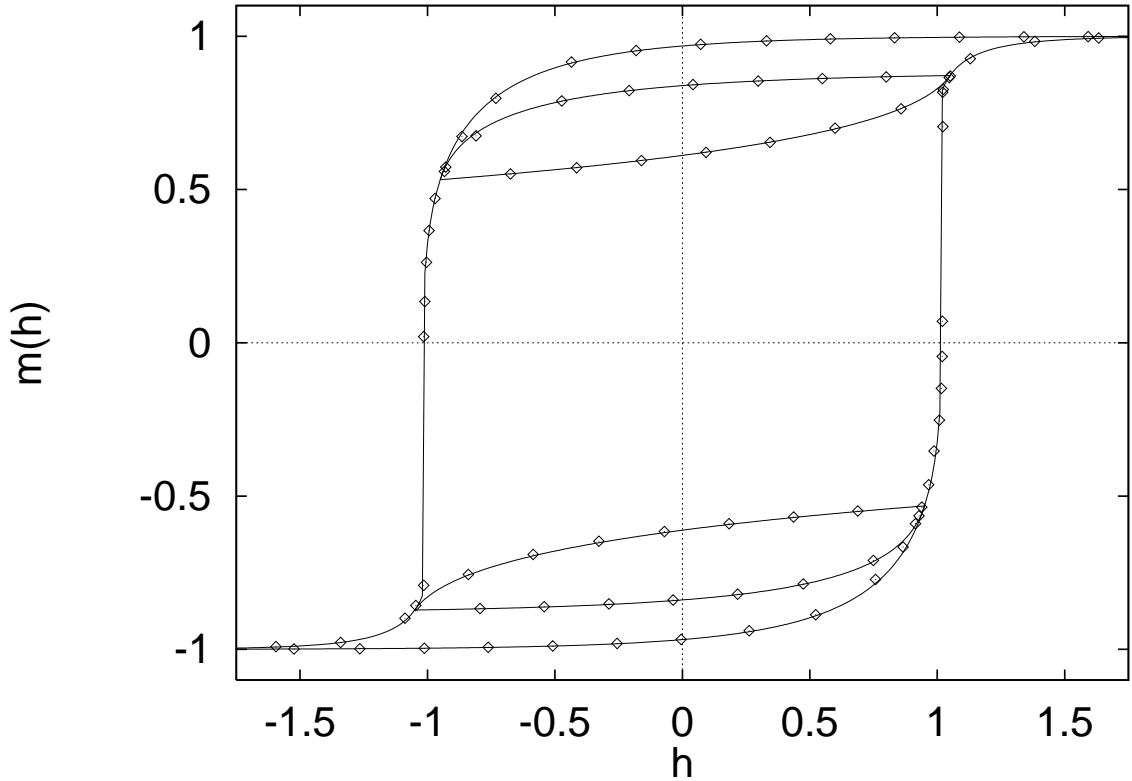}

\caption{ Hysteresis in the ferromagnetic RFIM on a Bethe lattice with
coordination number z=4, and a Gaussian distribution of random field with
mean zero and $\sigma = 1.7$. Continuous lines show the theoretical
result, and dots mark the results of numerical simulations.  Notice the
first order jumps in the magnetization at applied field $h \approx \pm 1$
(J=1) because $\sigma < \sigma_{c}$. Two minor loops within the major loop
are obtained by reversing the increasing field at $h=.95$, and $h=1.05$
respectively. The minor loops touch the upper half of the major loop when
the field has been reversed by an amount 2 $|J|$ (see text).}

\end{center}
\end{figure}

\begin{figure}
\begin{center}

\includegraphics[angle=-90,width=16cm]{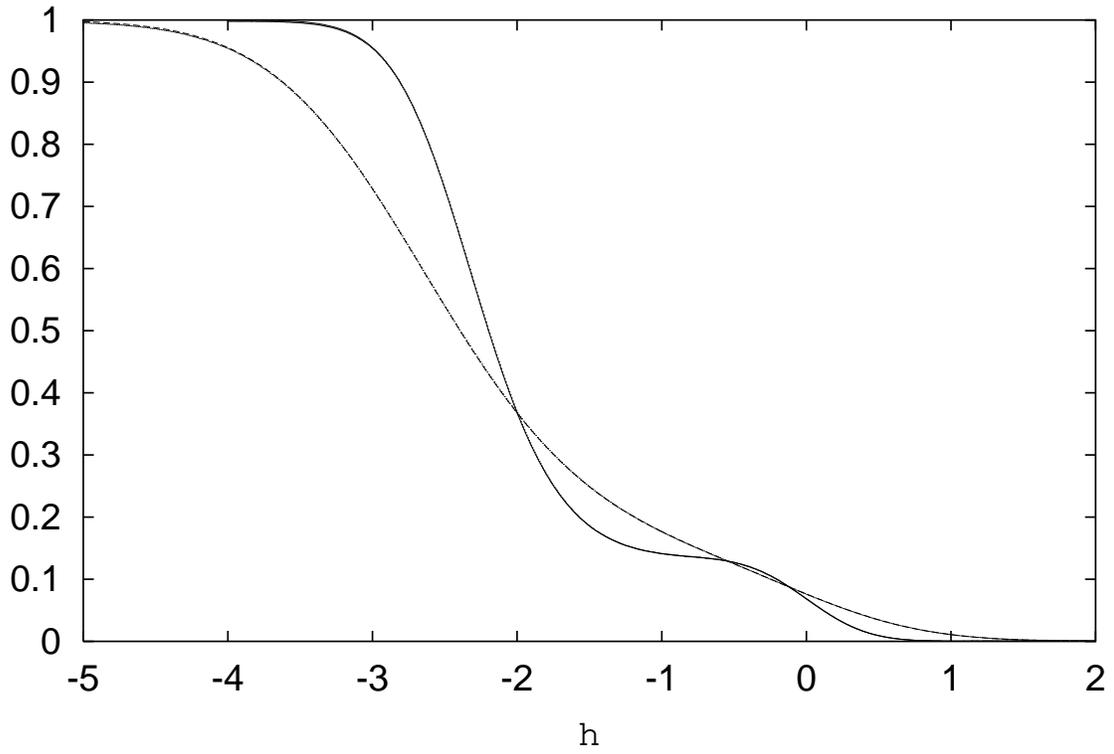}

\caption{ Occurrence of doublets in a one dimensional anti-ferromagnetic
RFIM. The y-axis shows the probability of occurrence of a doublet
$P_{\downarrow \downarrow}(h)$ at an applied field $h$. The curve with a
plateau like region in the middle is for $\sigma =.5 |J|$, and the other
curve is for $\sigma =|J|$. Results from numerical simulations have been
superimposed on the theoretical curves. }

\end{center}
\end{figure}


\newpage

\begin{thebibliography}{99}

\bibitem{sethna} J P Sethna, K A Dahmen, S Kartha, J A Krumhansl, B W
Roberts, and J D Shore, Phys Rev Lett 70, 3347 (1993).

\bibitem{shukla1}P Shukla, Physica A 233, 235 (1996).

\bibitem{dhar}D Dhar, P Shukla, and J P Sethna, J Phys A: Math. Gen.  
30, 5259 (1997).

\bibitem{sabha1} S Sabhapandit, P Shukla, and D Dhar, J Stat Phys 98, 103
(2000).

\bibitem{shukla2} Prabodh Shukla, Phys Rev E 62, 4725 (2000).

\bibitem{shukla3} Prabodh Shukla, Phys Rev E 63, 27102 (2001).

\bibitem{sabha2} S Sabhapandit, Ph D thesis, University of Mumbai (2002).
(2002).


\bibitem{shukla4} P Shukla, Physica A 233, 242 (1996); P Shukla, R Roy,
and E Ray, Physica A 275, 380 (2000); P Shukla, R Roy, and E Ray, Physica
A 276, 365 (2000).



\bibitem{sabha3} S Sabhapandit, D Dhar, and P Shukla, Phys Rev Lett 88,
197202 (2002).


\end{thebibliography}
\end{document}